\documentclass[
	twocolumn,
	amsmath,
	amssymb,
	prx,
	aps,
	nofootinbib,showpacs,superscriptaddress,
	]{revtex4-2}

\usepackage{graphicx}
\usepackage{amsmath}
\usepackage{amsfonts}
\usepackage[english]{babel}
\usepackage[normalem]{ulem} 

\usepackage[dvipsnames]{xcolor}

\renewcommand{\O}{\Omega}
\newcommand{\dO}{\dot{\Omega}}
\newcommand{\plane}{$(\O, \dO)$}

\begin{document}
\renewcommand{\figurename}{Fig.}

\title{Dynamic driving enables independent control of\\ material bits for targeted memory}

\author{E.~Gutierrez-Prieto}
\thanks{These authors contributed equally.}
\affiliation{Flexible Structures Laboratory, Institute of Mechanical Engineering, École Polytechnique Fédérale de Lausanne (EPFL), 1015 Lausanne, Switzerland}

\author{C.M.~Meulblok}
\thanks{These authors contributed equally.}
\affiliation{Huygens-Kamerlingh Onnes Lab, Universiteit Leiden, P.O.~Box~9504, NL-2300 RA Leiden, Netherlands}
\affiliation{AMOLF, Science Park 104, 1098 XG Amsterdam, Netherlands}

\author{M.~van~Hecke}
\affiliation{Huygens-Kamerlingh Onnes Lab, Universiteit Leiden, P.O.~Box~9504, NL-2300 RA Leiden, Netherlands}
\affiliation{AMOLF, Science Park 104, 1098 XG Amsterdam, Netherlands}

\author{P.M.~Reis}
\affiliation{Flexible Structures Laboratory, Institute of Mechanical Engineering, École Polytechnique Fédérale de Lausanne (EPFL), 1015 Lausanne, Switzerland}

\begin{abstract}
    Mechanical metamaterials with bistable elements can store vast amounts of information, but writing these memories requires impractical local control or lengthy multi-cycle protocols. We overcome this limitation with a dynamic control strategy that accesses any configuration in a single global drive cycle by leveraging the system's sensitivity to the drive and its time derivatives. We realize this strategy with bistable beams on a rotating platform, where drive cycles become orbits in a control space of angular velocity and acceleration. State changes occur when these orbits cross switching thresholds, which we rationally design so that each state can be accessed by a single drive orbit. We construct a five-bit system and demonstrate its full addressability by selecting drive orbits that write all 26 uppercase letters of the alphabet in ASCII representation. This dynamic control paradigm offers a general route towards smart, remotely operated devices across various physical domains.
\end{abstract}

\maketitle

\section{Introduction}
Mechanical structures made from bistable elements can store configurational memories~\cite{BensePNAS2021,LiuPNAS2024,KwakernaakPRL2023,MunganPRL2019,HeckePRE2021,KeimSciAdv2021,LindemanSciAdv2021,DingJCP2022,ShohatPNAS2022,MelanconAdvFuncMat2022,JulesPRR2022,LindemanSciAdv2025} to enable tunable stiffness~\cite{SilverbergSci2014,ChenNat2021,MirzajanzadehSciAdv2025}, sequential shape-morphing~\cite{FilipovPNAS2015,GuseinovNatComm2020,MeeussenNat2023}, and computation~\cite{McevoySci2015,KasparNat2021,YasudaNat2021,BensePNAS2021,ElHelouNat2022,KwakernaakPRL2023,SiroteNatComm2024,LiuPNAS2024,MeulblokArXiv2025}. While control over configurations is essential, it typically requires local and individual access to all elements~\cite{SilverbergSci2014,ChenNat2021,MirzajanzadehSciAdv2025}. Here, we introduce a dynamic control strategy that enables transitions between any pair of configurations within a single global drive cycle by exploiting sensitivity to different (orders of) derivatives of the driving. We illustrate this approach using bistable beams that snap through based on their collective rotation rate and acceleration. Through rational design and selection of drive cycles, we demonstrate targeted transitions along controllable pathways, including the writing of five-bit memories that encode all alphabetic characters. This form of dynamical control can be generalized to inertial, fluidic, electromagnetic, and electronic systems, thus providing a powerful method for writing memories and enabling smart, remotely controllable devices for applications in microfluidics, implants, smart infrastructure, and underwater or medical robotics.

Developing driving protocols that access all configurations of mechanical systems of material bits (mbits), comprising bistable elements, is essential for the functionality of multistable (meta)materials ~\cite{MunganPRL2019,HeckePRE2021,KeimSciAdv2021,LindemanSciAdv2021,DingJCP2022,ShohatPNAS2022,MelanconAdvFuncMat2022,JulesPRR2022,LindemanSciAdv2025,SilverbergSci2014,ChenNat2021,MirzajanzadehSciAdv2025,FilipovPNAS2015,GuseinovNatComm2020,MeeussenNat2023,McevoySci2015,KasparNat2021,YasudaNat2021,BensePNAS2021,ElHelouNat2022,KwakernaakPRL2023,SiroteNatComm2024,LiuPNAS2024,MeulblokArXiv2025}. Unfortunately, individually controlling each mbit becomes impractical as their number $N$ grows large~\cite{SilverbergSci2014,ChenNat2021,MirzajanzadehSciAdv2025}, and, even though it is possible to design mbits such that a single global driving field can access all states, such strategies require baroque driving protocols featuring up to $N$ drive cycles ~\cite{MunganPRL2019,TerziPRE2020}. However, there are several examples of physical systems that also respond to the driving rate: in mechanics, driving forces depend on the configuration and motion (inertia); in fluids, on static pressure and flow (viscosity); and in electronics, currents depend on voltage (resistance) and its rate of change (capacitance).
This poses the question of how and whether dynamic control strategies can be developed that allow access to all $2^N$ states using a single global drive cycle.

To address this challenge, we introduce a rational design principle to select dynamic driving protocols that together enable a global driving signal to steer transitions between arbitrary mbit states along selectable pathways.
As we show, at the core of our strategy lies the fact that dynamic driving can be naturally represented as driving orbits in a multi-dimensional space, where the different axes represent the relevant derivatives of the drive. The ability to connect regions in this space to different stable configurations -- unconstrained by the well-ordering of one-dimensional space that governs quasistatic driving -- is crucial for enabling any-to-any transitions and simplifies both system design and driving cycle selection.

\section{Phenomenology}
\begin{figure}[t!]
    \centering
    \includegraphics[width=0.5\textwidth]{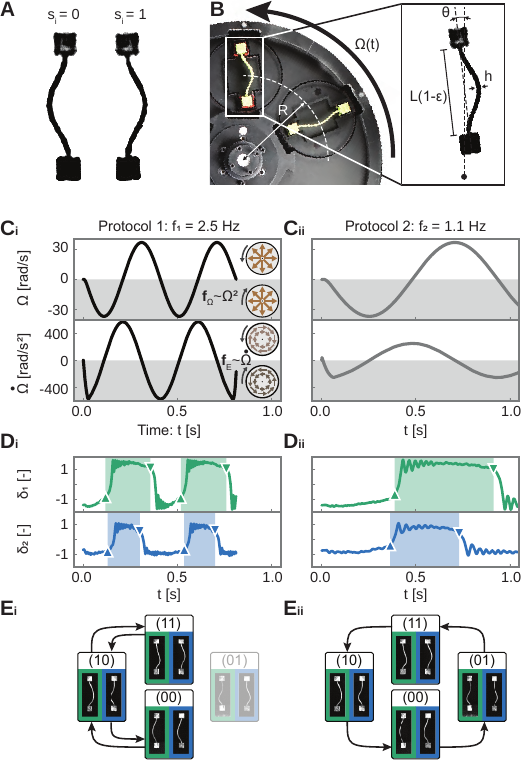}
    \caption{
    \textbf{Dynamic Driving of mbits.} 
    \textbf{A.}~Buckled beams form mbits whose buckling direction defines their state: $s_i = 0$ for counterclockwise and $s_i = 1$ for clockwise deflection
    \textbf{B.}~Rotating-disk apparatus showing system I and design parameters (see S.M.).
    \textbf{C.}~ Driving protocols
   $\Omega(t) = -a_j\sin(2\pi f_jt)$, where
    $a_1=a_2 = 36.5~\text{rad/s}$ and frequencies $f_i$ as indicated. 
    \textbf{D.} Evolution of beam midspan $\delta_i$ showing distinct switching sequences connecting $S: (00)\rightarrow(11)$ via either $(01)$ or $(10)$. 
    \textbf{E.} Corresponding pathways of states.
    }
    \label{fig:01}
\end{figure}

While the strategy introduced above should be general for a wide range of physical systems, we focus on a mechanical implementation where each mbit $i$ is a pre-buckled, double-clamped elastic beam mounted on a rotating disk, with its buckling direction defining the binary state $s_i$ (Fig.~\ref{fig:01}A; see S.M. and Movie S1)~\cite{KwakernaakPRL2023,GuerraPRL2023,MeulblokArXiv2025-02}. 
The angular velocity $\Omega(t)$ and time-derivative $\dot{\Omega}(t)$ control the radial centrifugal forces $f_\Omega \sim \Omega^2$ and azimuthal Euler forces $f_E \sim \dot{\Omega}$ that jointly control the switching of the mbits~\cite{GutierrezEML2024} (Fig.~\ref{fig:01}B). This interplay enables $\Omega(t)$ to serve as a dynamic control input for selectable transition pathways.

We demonstrate this selectivity with system I (Fig.~\ref{fig:01} and S.M. Movie S2), which comprises two mbits that differ in compressive strain $\varepsilon$ and tilt angle $\theta$ (see S.M., table~\ref{tab:systems}). We consider the transition pathways under two dynamic (harmonic) driving protocols $\Omega(t) =- a_j\sin(2\pi f_jt)$, where $a_j$ is fixed and $f_j$ is varied, to induce transitions between collective states $S=(s_1s_2)=(00)$ and $(11)$. 
Strikingly, while for $f_1$ the system goes through intermediate state $(10)$, for $f_2$ the pathway from $(00)$ to $(11)$ exhibits a different intermediate state $(01)$ (Fig.~\ref{fig:01}E,H; S.M. Movie S2). 
To understand how dynamic driving shapes transition pathways, we map driving protocols to orbits in the $\Omega$–$\dot\Omega$ plane (Fig.~\ref{fig:02}A, S.M. Movie S2). Each protocol traces a distinct elliptical orbit, along which we plot the observed states and the up ($s_i\!:0\!\rightarrow\!1$) and down ($s_i\!:1\!\rightarrow\!0$) transitions of both mbits, revealing that switching thresholds depend on both $\Omega$ and $\dot\Omega$
(Fig.~\ref{fig:01}C--H; Fig.~\ref{fig:02}A). A central insight is that transitions occur along \emph{up} curves $u_i$ ($s_i\!:0 \rightarrow 1$) and \emph{down} curves $d_i$ ($s_i\!:1 \rightarrow 0$) in the $\Omega$--$\dot\Omega$ plane, which can be identified by aggregating data from many elliptical driving orbits (Fig.~\ref{fig:02}A; see S.M.). The up and down transition curves for each element partition the plane into domains corresponding to distinct stable states. Note that all states are stable at the origin (\textit{i.e.}, without driving), and that only crossings oriented away from the origin can produce transitions (Fig.~\ref{fig:02}A). Intersections of transition curves are key; for example, the two elliptical driving orbits cross $u_1$ and $u_2$ on opposite sides of their 
intersection, thus producing the distinct pathways: $(00) \!\rightarrow\! (10) \!\rightarrow\! (11)$ and $(00) \!\rightarrow\! (01) \!\rightarrow\! (11)$ (Fig.~\ref{fig:02}A and S.M. Movie S2).

\section{Orbit selection}
\begin{figure*}[t!]
   \centering
   \includegraphics[width=1\textwidth]{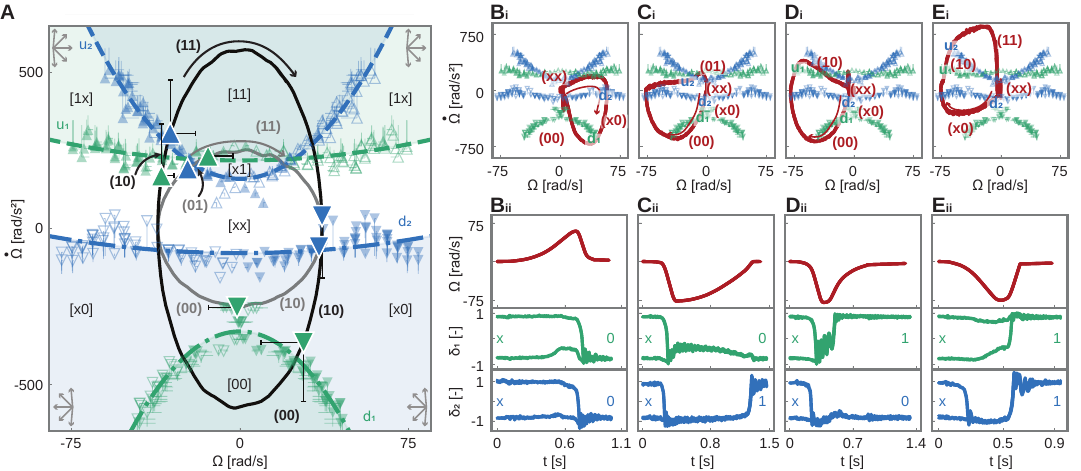}
   \caption{
    \textbf{Orbits, transitions and states.}
    \textbf{A.}~Data for System I in the \plane-plane, where green and blue correspond to mbits one and two, respectively. The ellipses correspond to protocol 1 (black) and 2 (grey), with arrows indicating the required clockwise orientation. The observed up transitions ($\triangle$), down transitions (\rotatebox[origin=c]{180}{$\triangle$}), 
    and states highlight the difference in intermediate states (bold).
    The switching onset for harmonic protocols with (frequency $f_j=1.5\,\mathrm{Hz}$, drive amplitude $a_j\in[5.0,\,80.0]\,\mathrm{rad/s}$) are shown as faint symbols with associated uncertainties (see S.M.). The dashed and dot-dashed lines are quadratic fits serving as guides to the eye for the up- and down-switching curves, respectively (see S.M.). These lead to partitions (shaded regions) where one or more collective states (square brackets) are stable, where `x' indicates mbits that can be both 0 and 1. The arrows indicate the allowed orbit directions in each quadrant (open intervals of size $\pi$).
    \textbf{Bi-Ei.}~Selected orbits that write specific targeted states: 
    \textbf{B}~(00), 
    \textbf{C}~(01), 
    \textbf{D}~(10) and 
    \textbf{E}~(11).
    \textbf{Bii-Eii.}~Angular velocity $\Omega(t)$ corresponding to selected orbits (top panels) and corresponding evolution of the midspan displacement of beams one (middle panels) and two (bottom panels), starting from any of four initial states, demonstrating how the final state is independent of the initial configuration.}\label{fig:02}
\end{figure*}

We leverage the principles elucidated above to selected driving orbits for system I so that any of its four binary states can be written (Fig.~\ref{fig:02}B--E and S.M. Movie S3). We focus on orbits that start and end at the origin, where all states are stable. Note that any orbit with appropriate orientation (Fig.~\ref{fig:02}A)
can be mapped to a physically realizable input function $\Omega(t)$  (see S.M.). Reaching a target state requires changing mbits from $0 \!\rightarrow\! 1$ or  $1 \!\rightarrow\! 0$, which demands appropriate crossings of the corresponding $u_i$ and $d_i$ curves. For example, an orbit that crosses only $d_1$ and $d_2$ zeroes both mbits and produces $S=(00)$, regardless of the initial state (Fig.~\ref{fig:02}B).  Note that specific switching curve
topologies complicate orbit selection. Naively, one might expect that the target state $S=(01)$ can be written by simply crossing
$d_1$ and $u_2$, but any such orbit also crosses $d_2$ (Fig.~\ref{fig:02}C). In system I, writing the target state $S=(01)$ requires an orbit that crosses $d_2$, $d_1$ and then $u_2$ (Fig.~\ref{fig:02}C). Similarly, all other target states of system I can be reached by selecting appropriate orbits (Fig.~\ref{fig:02}D,E). These examples demonstrate that dynamic control enables the rational selection of targeted orbits capable of writing any combination of mbits in system I.

\begin{figure*}[hbt]
    \centering
    \includegraphics[width=1\textwidth]{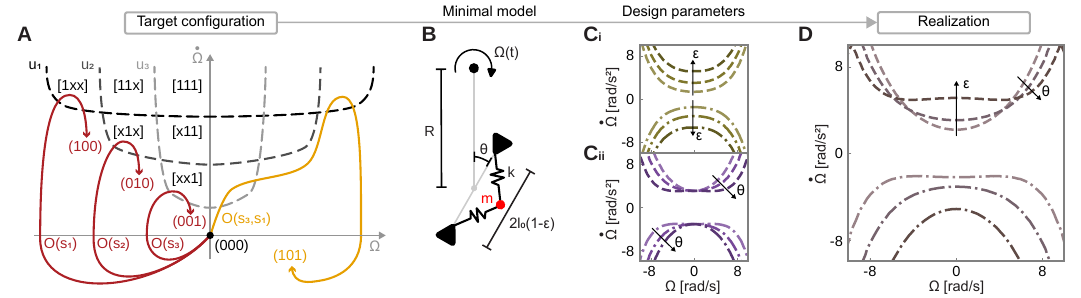}
    \caption{
    \textbf{System design.}
    \textbf{A.} 
    Schematic set of hierarchical switching curves $(u_1, u_2, u_3)$ with selected orbits that enable the writing of arbitrary states: starting from state $(000)$, the orbits $O(s_1)$, $O(s_2)$, $O(s_3)$ and $O(s_3,s_1)$ write states $(100)$, $(010)$, $(001)$ and $(101)$.
    \textbf{B.} Minimal model for mbits, comprising two linear springs (spring constant $k$ and rest length $\ell_0$), each connected to a central mass $m$;  the key design parameters are the compressive strain $\varepsilon$ and tilt angle $\theta$.
    \textbf{C.} Representative trends for the transition curves obtained from the minimal model with parameters $\varepsilon$ and $\theta$ (arrows). Top: $\varepsilon = (0.03,\,0.05,\,0.07)$ with $\theta=0$; Bottom:  
    $\theta = (-3^\circ,\,0^\circ,\,3^\circ)$ with $\varepsilon=0.05$ . 
    Experimental curves show similar trends (see S.M. Fig.~\ref{fig:BC_effect}).
    \textbf{D.} Hierarchical up transition curves for $(\varepsilon, \theta) = [(0.04, -3^\circ), (0.05, 0^\circ), (0.07, 4^\circ)]$.
    }\label{fig:03}
\end{figure*}

\begin{figure*}[bht]
    \centering
    \includegraphics[width=1\textwidth]{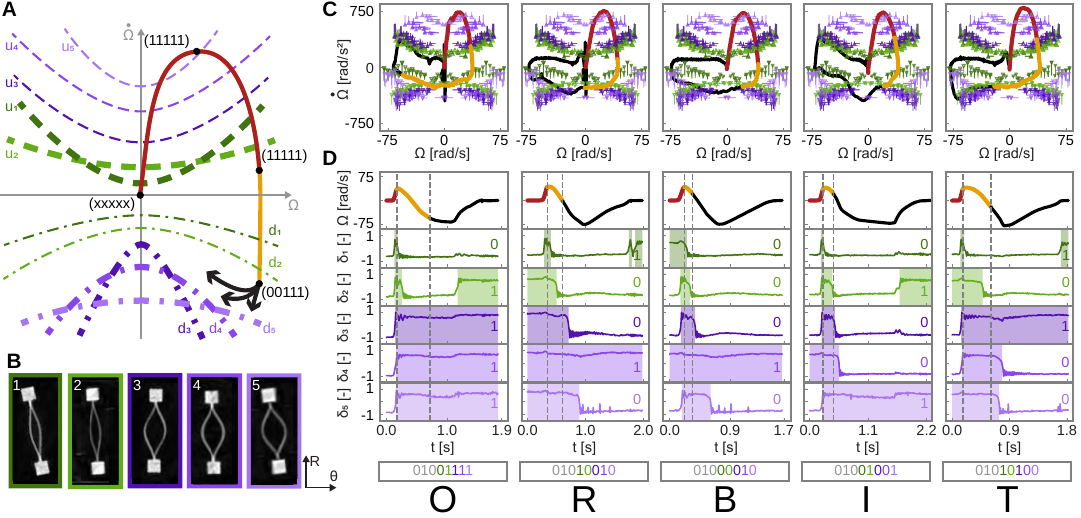}
    \caption{\textbf{Writing arbitrary states in System II.}
    \textbf{A.} Target hierarchical organization of the curves $(u_1,u_2)$ and $(d_3,d_4,d_5)$ (bold), together with appropriate placement of the other switching curves. Starting out from arbitrary state $(\mathrm{xxxxx})$, the first phase of the erase orbit (red) sets all bits, reaching state $(11111)$, whereas the second phase (gold) resets the first two bits reaching state $(00111)$, after which the writing phase commences (black).
    \textbf{B.}  System II is comprised of five mbits with rationally designed tilt angle and compression that lead to the targeted ordering of switching curves.
    \textbf{C.} Driving orbits corresponding to specific characters (red: erase phase one, gold: erase phase two, black: writing phase). 
    \textbf{D.} Experimental driving protocol and beam evolution demonstrate how dynamic control enables writing ``ORBIT'' (see S.M. Movie S4). The ASCII encoding standard is used to map the bit strings to the Latin characters~\cite{ASCII}. Although ASCII requires eight bits, the first three bits are identical for all uppercase characters; hence, we write the remaining five bits. Below each orbit, the full bit string is shown, with colors labeling the mbits, and gray indicating the three shared bits.  
    }\label{fig:04}
\end{figure*}

To be able to write any possible state in a system with an arbitrary number of mbits, we first introduce hierarchically organized switching curves as a design target and then show how the physical mbits can be designed to realize such curves. As a schematic example, consider three up curves that intersect such that there is a range in driving space where each is closest to the origin (Fig.~\ref{fig:03}A). Initializing the system in state $(000)$, it is then possible to select driving orbits $O(s_i)$ that set mbit $i$ to 1 (Fig.~\ref{fig:03}A). Setting multiple mbits to 1 can be achieved by chaining such orbits or by selecting more complex multi-crossing driving orbits such as $O(s_3,s_1)$; this freedom allows control of intermediate states (Fig.~\ref{fig:03}A). Our proposed hierarchical design can be generalized to encompass more mbits, and is the prime design target to realize systems with highly targeted switching behavior, as we demonstrate next.

\section{System design}
We explore the real-space design of the mbits by investigating the dependence of the switching curve on the tilt angle $\theta_i$ and the compressive strain $\varepsilon_i$ of the corresponding mbits. To guide us with a rational design, we use a von~Mises-truss model comprising a mass connected to two linear springs (see S.M.). This model system captures the bistability of the mbits and, when subjected to centrifugal and Euler forces, allows
to study the properties of the switching curves as a function of 
$\theta_i$ and $\varepsilon_i$ ~\cite{MisesZAMM1923,BelliniJNLM1972,Cedolin2010} (Fig.~\ref{fig:03}B).
We first fix $\theta=0$ and find that the switching curves can be moved away from the origin by increasing $\varepsilon$, which controls the energy barrier between the $s=0$ and $s=1$ states (Fig.~\ref{fig:03}C$_\mathrm{i}$). Second, non-zero tilt angles ($\theta\ne0$) break the symmetry between the up and down switching curves, allowing widening of the $u$ curves and narrowing of the $d$ curves by increasing $\theta$ (Fig.~\ref{fig:03}C$_\mathrm{ii}$).
Systematically varying the design parameters ($\theta_i$ and $\varepsilon_i$) enables the construction of the hierarchical switching curves: \emph{up-curves} require increasing strain and tilt, while \emph{down-curves} combine increasing strain with decreasing tilt (Fig.~\ref{fig:03}D). Hence, geometric control of the mbits facilitates designing systems with targeted hierarchical switching curves.

\section{Sequential memory writing}
Finally, we design a five-mbits system, whose collective state can be written with a single clockwise driving orbit starting and terminating at 
$(\Omega,\dot{\Omega})=(0,0)$. As a first step, we design the switching curves such that a target state $(t_1 t_2 t_3 t_4 t_5)$ can be reached from an initial arbitrary state $(\mathrm{xxxxx})$. Hierarchical design of either all up or all down switching curves is sufficient to do so in principle, but, instead, to enhance robustness, it is advantageous to leverage hierarchical ordering of subsets of the up and down curves, so that less critical intersections are required. Here, we opt for a design where $(u_1,u_2)$ and $(d_3,d_4,d_5)$ are hierarchically ordered, but note that other hierarchical orderings are valid too. 
We then use orbits that consist of a first erase phase, where the state evolves to $(00111)$ via the intermediate state $(11111)$, and, then, a write phase, where the first two bits can be switched $0 \!\rightarrow\! 1$
and the last three bits can be switched $1 \!\rightarrow\! 0$ (Fig.~\ref{fig:04}A). In such a scenario, the precise locations of the non-hierarchical curves are less critical, provided that $(u_1,u_2)$ can be crossed without crossing $(u_3,u_4,u_5)$, and $(d_1,d_2)$ are crossed before crossing $(d_3,d_4,d_5)$, so that in the erase phase we can reach state $(00111)$ (Fig.~\ref{fig:04}A). Using our numerical toolbox and some experimental fine-tuning, we realize the target ordering of the switching curves experimentally in System II (Fig.~\ref{fig:04}B; see S.M.).  

To demonstrate the expressive capabilities of our approach, we construct a library of driving orbits capable of writing all 26 binary states $S$ that encode the Latin upper-case characters of the alphabet (see S.M. Fig.~\ref{fig:alphabet}). Selecting appropriate orbits, as an example, we sequentially drive our five-mbit system to write the word ``ORBIT'' (Fig.~\ref{fig:04}C--D; see S.M. and Movie S4). This capability of writing arbitrary words using Latin characters highlights that dynamic control of a rationally designed mbit system
offers a powerful strategy to write and erase information.

\section{Discussion}
We stress that dynamic driving extends beyond rotation, encompassing inertial or viscous effects in mechanics \cite{DykstraJAM2019,DykstraAPLM2022,JulesCommPhys2023,LindemanPRL2023,jinArXiv2025}, and, more broadly, dynamic effects in fluid flow \cite{GeddesPRE2010,MartinezCalvoNatComm2024}, chemical reactions \cite{vanSluijsNatComm2022}, and electronic circuits \cite{AltmanArXiv2025}.
Their systematic explorations will unlock agile information encoding for future smart and multifunctional devices.

\section{Methods}
\label{sec:exp_methods}
\subsection{Experimental Methodology}

\noindent \textbf{Beams:} 
The beams are fabricated by casting  a silicone-based elastomer (VPS32, Zhermack Elite Double 32, with Young's modulus $E = 1.22 \pm 0.05\,\mathrm{MPa}$, Poisson’s ratio $\nu\approx 0.5$, and density $\rho = (1.17 \pm 0.01)\cdot 10^3\, \mathrm{kg}\, \mathrm{m}^{-3}$~\cite{GrandgeorgeJMPS2022,LeroyPRF2022}) 
using a fully enclosed mold. Each mold consists of multiple layers of laser-cut acrylic components to produce uniform rectangular beams with thickness $h \in [1.5,\,3.0]\,\mathrm{mm}$, length $L = 80\,\mathrm{mm}$, and width $b = 8\,\mathrm{mm}$. The extremities of the beams are cast to blocks with a larger width ($b\times h\times l=15\times12\times15~\mathrm{mm}$) to emulate clamped boundary conditions.\\

\noindent \textbf{Placement:} 
Each beam is aligned such that its width points out of the rotary plane and is placed in a laser-cut acrylic holder that can set the compressive strain in the range  $\varepsilon\in[0.01,\,0.1]$ with a resolution of $\pm0.001$. These holders are mounted on a rigid disk in a way that the tilt angle can be varied in the range $\theta\in[-15^\circ,\,15^\circ]$ with a resolution of $\pm0.5^\circ$. Finally, the holders are enclosed by a transparent acrylic plate to minimize air drag during rotation (see Fig.~\ref{fig:01} and Supplementary Movie 1). \\

\noindent\textbf{System parameters:} In Table~\ref{tab:systems}, we present the geometric parameters of the mbits used in the two experimental systems (I and II) investigated in the main text. 

\begin{table}[htb]
\centering
\begin{tabular}{|l||c|c||c|c|c|c|c|}
\hline
                        & \multicolumn{2}{c||}{System I} & \multicolumn{5}{c|}{System II} \\ \hline
mbit                   & 1   & 2  & 1 & 2 & 3 & 4 & 5 \\ \hline \hline
$L$ [mm] & $80$ & $80$ & $80$ & $80$ & $80$ & $80$ & $80$ \\ \hline
$h$ [mm] & $1.9$ & $1.9$ & $1.6$ & $1.9$ & $2.1$ & $2.3$ & $2.6$ \\ \hline
$b$ [mm] & $8.0$ & $8.0$ & $8.0$ & $8.0$ & $8.0$ & $8.0$ & $8.0$ \\ \hline
$R$ [mm] & 150 & 150 & 150 & 150 & 150 & 150 & 150 \\ \hline
$\varepsilon$ [-] & $0.050$ & $0.015$ & $0.015$ & $0.026$ & $0.037$ & $0.055$ & $0.060$ \\ \hline
$\theta$ [$^\circ$]& $-4.5$  & $4.5$  & $4.5$  & $1.5$  & $0.0$ & $4.5$  & $7.5$ \\ \hline
\end{tabular}
\caption{\textbf{Parameters of the mbits in systems I and II}. The length $L$, in-plane thickness $h$, and out-of-plane width $b$ have an experimental uncertainty of $\pm0.1~\text{mm}$. The radial position $R$ of the beams has an uncertainty of $\pm1~\text{mm}$. The uncertainties in compression $\varepsilon$ and angle $\theta$ are, respectively, $\pm0.001$ and $\pm0.5^\circ$.}\label{tab:systems}
\end{table}

\noindent \textbf{Rotational apparatus:} 
The rigid disk holding all beams is mounted to a torque motor (ETEL RTMBi140-050A) capable of delivering up to $122~\mathrm{Nm}$ of peak torque. The motor is driven by a Modular AccurET 600V/40A controller, which we program with a time series of angular positions to generate the desired velocity profile $\Omega(t)$. This setup enables precise trajectory execution, with maximum angular velocity and acceleration of $\pm 85~\mathrm{rad\cdot s^{-1}}$ and $\pm 10^3~\mathrm{rad\cdot s^{-2}}$, respectively. An encoder records the disk's angular position at $20~\mathrm{kHz}$, with an uncertainty of $\pm8\cdot10^{-4}~\mathrm{rad}$. 
To determine the angular velocity $\O(t)$ and acceleration $\dO(t)$, we apply a third-order Savitzky--Golay filter (15-point window) to the position data and perform numerical differentiation using fourth-order central finite differences.\\

\noindent \textbf{Image analysis:} The real-space configuration of each beam is imaged using a global shutter camera (IDS U3-3040SE-M-GL, with lens Vision-Lens 3.5mm EFL, F/1.4) is placed in the lab frame, above the rotating disk and aligned with its center of rotation. The shutter time is set to 100$\mu s$ to minimize image blur during rotation. We set a region of interest of $346{\times}346$ pixels, with 1.2~mm/pixel, to image the entire disk, setting a frame rate of 743~fps.

To quantify the configuration and dynamics of the beams, we developed an in-house image analysis algorithm with support from the EPFL Imaging Hub (Dr. Florian Aymanns). First, we locate two white markers, one at the center of rotation and the other at $R=50~\mathrm{mm}$ from the center of rotation (see Extended Data Fig.~\ref{fig:im_prc}). Next, for every frame and all measured beams, the image analysis involves several key steps (see Extended Data Fig.~\ref{fig:im_prc}b): 
(1) We detect the two white markers to establish the angular position, 
$\varphi$, of the disk, and find that it is in good agreement with the output of the encoder.
(2) A neural network, trained by manually labeling clamp corners, identifies the corners of each clamp. 
(3) To approximate the neutral line of each beam, we use Hermite exponential splines from the Python library \texttt{splinebox}~\cite{splinebox}. 
(4) We strategically place five spline knots: two at the identified beam edges and three additional knots equally distributed along the beam's longitudinal axis. 
(5) To precisely locate these central knots, we define three equally spaced lines normal to the clamp-to-clamp line. 
(6) Finally, on each line, we identify the point of maximum brightness by fitting a Gaussian function to the brightness distribution along the line. We locate the knots at the maximum of the fitted Gaussian.
This image processing method achieves sub-pixel resolution (with $\pm0.2\,\mathrm{mm}$ uncertainty) in the midpoint deflection of the centerline $\delta$ and generates a time series of the centerline of each beam represented in splines (see Extended Data Fig.~\ref{fig:im_prc}c).\\

\noindent \textbf{Definition of snap-through events:} To identify the snap-through of each beam, we start from the detected midpoint deflection $\delta(t)$ and follow a four-step approach:
(1) We identify the times at which the midpoint of beam $i$ crosses its longitudinal axis $\delta_i(t_{i}^0)=0$ (orange crosses, Extended Data Fig.~\ref{fig:SI_snap}a-b). 
(2) We define the temporal window $W_i$ centered about $t_{i}^0$, $W_i = [t_{i}^0-25\,\mathrm{ms},t_{i}^0+25\,\mathrm{ms}]$,
and determine the maximum absolute velocity $|\dot{\delta}_i|_{\text{max}}$ achieved within $W_i$
(shaded area, Extended Data Fig.~\ref{fig:SI_snap}a).
(3) We characterize the time of switching-onset $t^*_i$ as $|\dot{\delta}_i(t^*_i)|= 0.15\,|\dot{\delta}_i|_{\text{max}}$ (purple triangles, Extended Data Fig.~\ref{fig:SI_snap}a). The threshold value 0.15 was empirically optimized to provide reliable switching-onset detection across our experiments.
(4) We specify ($\O(t^*_i)$, $\dO(t^*_i)$) as the switching-onset in the \plane-plane. We take ($\O(t^*_i)-\O(t_i^0)$, $\dO(t^*_i)-\dO(t_i^0)$) as an estimate of the uncertainty of the switching onsets, and indicate these with faint lines.
\\

\noindent \textbf{Switching curves:} We experimentally map the switching curves of each mbit by applying a set of harmonic driving protocols $\O(t) = a_j \sin(2\pi f_j t + \phi)$, where we fix the
frequency at $f=1.5\,\mathrm{Hz}$, and systematically vary the amplitude $a_j\in[5.0,\,80.0]\,\mathrm{rad}/\mathrm{s}$ with steps of $5.0\,\mathrm{rad}/\mathrm{s}$, and the phase $\phi = \{0,\,\pi\}\,\mathrm{rad}$ (see Extended Data Fig.~\ref{fig:SI_snap}b-c). 
Prior to each driving protocol, we set all mbits to $s_i=0$ for $\phi=0$ and to $s_i=1$ for $\phi=\pi$. This collection of driving protocols enables a precise characterization of the transition curves within the $\O/\dO < 0$ domain.
To complete the transition curves across the full domain, the experimental data is mirrored about $\O=0$; this symmetry operation is justified since the centrifugal force $f_\Omega\sim\O^2$ is independent of the sign of the velocity. As a guide to the eye for the switching-curves data in Fig.~\ref{fig:02}a, we fit a quadratic polynomial function to the measured snap-through points. 
To ensure a mirror-symmetry about $\O=0$, we consider a function without odd terms: $p(\O)=c_0+c_2\O^2$ (dashed and dot-dashed curves in Extended Data Fig.~\ref{fig:SI_snap}c, Fig.~\ref{fig:02}a). We note that although the fitted curves give a good estimation of the location of the switching curves, selected orbits near these curves may trigger switching.
Nevertheless, the selected orbits presented in this paper were repeated five times and consistently demonstrated the reported switching behavior.
\\


\subsection{Reduced-order model}

\noindent \textbf{Computational spring model:}
In parallel with the experiments, we use a reduced-order model to guide the rational design of the mbits switching curves. This model has a configuration analogous to a von~Mises truss~\cite{MisesZAMM1923,BelliniJNLM1972,Cedolin2010},
comprising a central point (mass $m$, position $\mathbf{r}$)
connected via two linear springs (spring constant $k$, rest length $l_0$) with end points $\mathbf{r_{s1}}$ and $\mathbf{r_{s2}}$ (Fig.~3b). We take
$\mathbf{r_{s1}}=(1-\varepsilon)(
\sin(\theta), \cos(\theta))$, and $\mathbf{r_{s2}}=-\mathbf{r_{s1}}$, to ensure bistability for $\varepsilon>0$.
Similar to experiments, the tilt angle $\theta$ and compressive strain $\varepsilon$ are the key design parameters. A total of four forces act on the central  mass: First, the two springs give rise to the two forces  $\mathbf{F_{si}}=k(||\mathbf{\Delta r_{si}}||-l_0)\mathbf{\Delta r_{si}}/||\mathbf{\Delta r_{si}}||$,
where 
 $\mathbf{\Delta r_{si}}=\mathbf{r} - \mathbf{r_{si}}$.
Second, rotation gives rise to the centrifugal force $\mathbf{F}_{\Omega} = -m \, \mathbf{\O} \times \left( \mathbf{\O} \times (\mathbf{r} +R\textbf{e}_y)\right)$, and Euler force $\mathbf{F}_{E} = -m \, \mathbf{\dO} \times (\mathbf{r} +R\textbf{e}_y)$, where 
$R$ is the distance between the center of rotation and the position of the point mass $\mathbf{r}$ at rest, when $(\O,\dO)=(0,0)$. We determine the equilibrium position(s) of the mass for a given angular velocity $\O$ and acceleration $\dO$ by numerically solving the force balance using gradient descent. To find the transition curve of the reduced-order model, we determine the monostable and bistable regions of each truss.\\

\subsection{Orbit selection}

\noindent \textbf{Realizing selected orbits:}
A target orbit can be described as
\((\Omega(\tau), \dot{\Omega}(\tau))\),
where $\tau$ is a pseudo-time. 
Such an orbit can be realized if and only if one can find a 
mapping between physical time $t$ and $\tau$, such that 
\begin{equation} \dot{\Omega}(\tau) = \text{d}\Omega(\tau)/\text{d}t = 
\left(\text{d}\Omega(\tau)/\text{d}\tau\right)\cdot (\text{d}\tau/\text{d}t)~.
\label{eq:cons}
\end{equation}
Since $\tau$ is a monotonically increasing
function of the physical time $t$, this 
equation implies 
the \textit{directionality constraint} on each orbit,
i.e., $\text{d}\Omega(\tau)/\text{d}\tau$ has the same sign as $\dot{\Omega}(\tau)$. To construct a specific orbit, we start from a given diskrete sequence of points in the \plane-plane, i.e., $(\Omega(\tau), \dot{\Omega}(\tau))$, and define
$\tau$ as the trajectory index $\tau=(0,1,2,\dots)\cdot \Delta\tau$. We then
use Eq.~(\ref{eq:cons}) to find $\Delta t(\tau)$:
\begin{equation}
{\Delta t(\tau) }= \frac{\Omega(\tau + \Delta \tau) - \Omega(\tau)}{\dot{\Omega}(\tau)} ~,
\label{eq:deltat}
\end{equation}
and map the trajectory index $\tau$ to real time  as
$t(\tau) = \sum_{0}^{\tau} \Delta t(\tau)$.
This procedure yields a time-parametrized angular velocity profile \(\Omega(t)\) that matches the selected orbit in \((\Omega(\tau), \dot{\Omega}(\tau))\)-space and can be used as input for angular velocity-controlled experiments.\\

\noindent \textbf{Acknowledgments} The authors acknowledge the EPFL Center for Imaging, especially Dr. Florian Aymanns, for assistance/help with developing the image processing pipeline.
C.M.M. and M.v.H. acknowledge funding from European Research Council Grant ERC-101019474.





\bibliographystyle{apsrev4-2}
\bibliography{ProjectDynamic}

\section*{Supplementary Figures}
\renewcommand{\figurename}{Supplementary Fig.}
\setcounter{figure}{0}

\begin{figure*}[h!]
    \centering
    \includegraphics[width=1\textwidth]{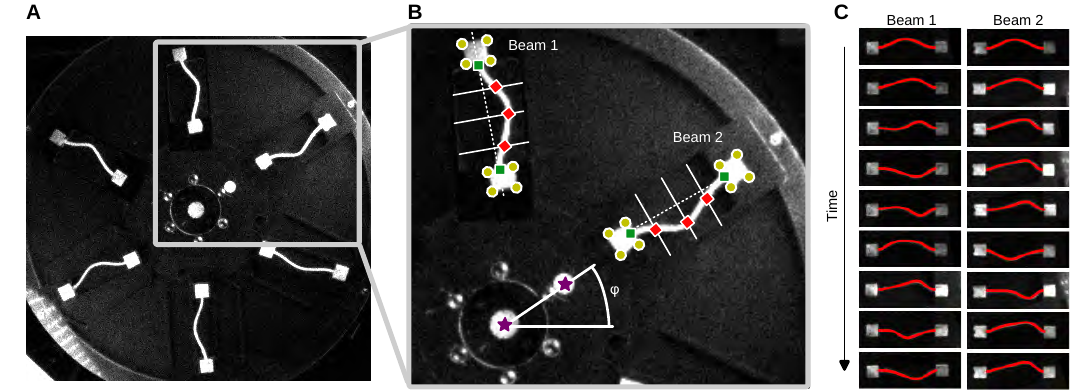}
    \caption{\textbf{Automated image‐processing pipeline.}
        \textbf{A.} Snapshot of the rotating device containing six flexible beams. 
        \textbf{B.} Schematic of the detection and spline‐fitting workflow. White circles mark fiducial markers whose centroids (stars) define the instantaneous rotation angle $\varphi$. A neural network locates the clamp corners (light green circles), from which the two outer nodes (dark green squares) are established to determine the beam’s neutral axis (dashed line). Three search lines, drawn perpendicular to this axis, detect intensity maxima corresponding to the remaining nodes used for cubic‐spline interpolation (red diamonds). 
        \textbf{C.} Time‐sequence montage of the fitted splines (red solid lines) overlaid on raw beam images (white), demonstrating sub‐pixel accuracy of image analysis.}
    \label{fig:im_prc}
\end{figure*} 

\newpage

\begin{figure*}[h!]
    \centering
    \includegraphics[width=1\textwidth]{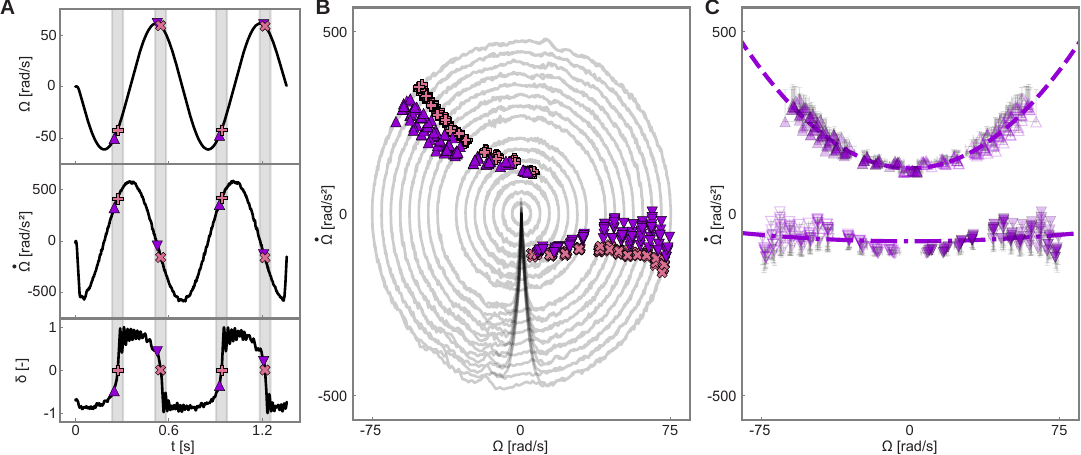}
    \caption{\textbf{Switching-curve extraction from snap-through dynamics.}
        \textbf{A.} From top to bottom: angular velocity $\Omega$, angular acceleration $\dot{\Omega}$, and midspan deflection $\delta$ of beam two from system I subject to harmonic driving protocol $\O(t) = a_j \sin(2\pi f_j t + \phi)$ with  $a_j=63.9\,\mathrm{rad}/\mathrm{s}$, $f=1.5\,\mathrm{Hz}$ and $\phi = \pi\,\mathrm{rad}$.
        The orange crosses and purple triangles represent the point at which $\delta=0$ and the onset of switching, respectively (see text).
        The shaded region depict $W_i$ (see S.M.).
        \textbf{B.} all measured responses harmonic driving protocols with $a \in [5.0, 80.0]\,\mathrm{rad\,s^{-1}}$. The gray circles indicate the driving protocols.
        \textbf{C.} The switching events represented with uncertainty (see S.M.). The solid and hollow symbols indicate the measured and mirrored data, respectively (see S.M.). The dashed and dot-dashed lines are the quadratic polynomial fits estimating the up- and down-curves, respectively.}
    \label{fig:SI_snap}
\end{figure*}

\newpage

\begin{figure*}[h!]
    \centering
    \includegraphics[width=1\textwidth]{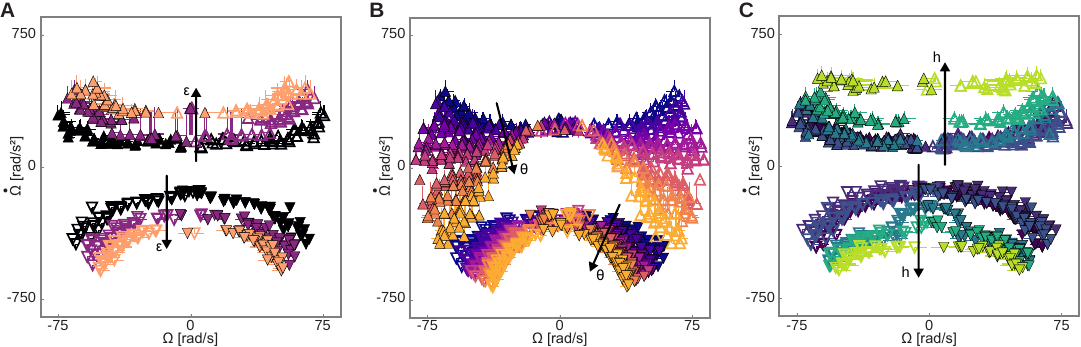}
    \caption{\textbf{Experimental mapping of transition curves across geometric parameters.}
        Each panel shows the observed up transitions ($\triangle$) and down transitions (\rotatebox[origin=c]{180}{$\triangle$}) as a function of one of the design parameters.
        \textbf{A.} 
        Strain $\varepsilon\in\{0.013, 0.037, 0.062\}$, with fixed tilt angle $\theta=1.5^\circ$ and beam thickness $h=1.9\,\mathrm{mm}$.
        \textbf{B.}
        tilt angle $\theta = \{1.5,\, 3.0^,\, 4.5,\, 6.0,\, 7.5,\, 9.0,\, 10.5,\, 12.0\}^\circ$, with fixed strain $\varepsilon=0.037$ and beam thickness $h=1.9\,\mathrm{mm}$.
        \textbf{C.}
        Beam thickness $h=\{1.6,\, 1.8,\, 1.9,\, 2.1,\, 2.4,\, 2.7\}\,\mathrm{mm}$, with fixed strain $\varepsilon=0.037$ and tilt angle $\theta=1.5^\circ$.}
    \label{fig:BC_effect}
\end{figure*}

\newpage

\begin{figure*}[h!]
    \centering
    \includegraphics[width=1\textwidth]{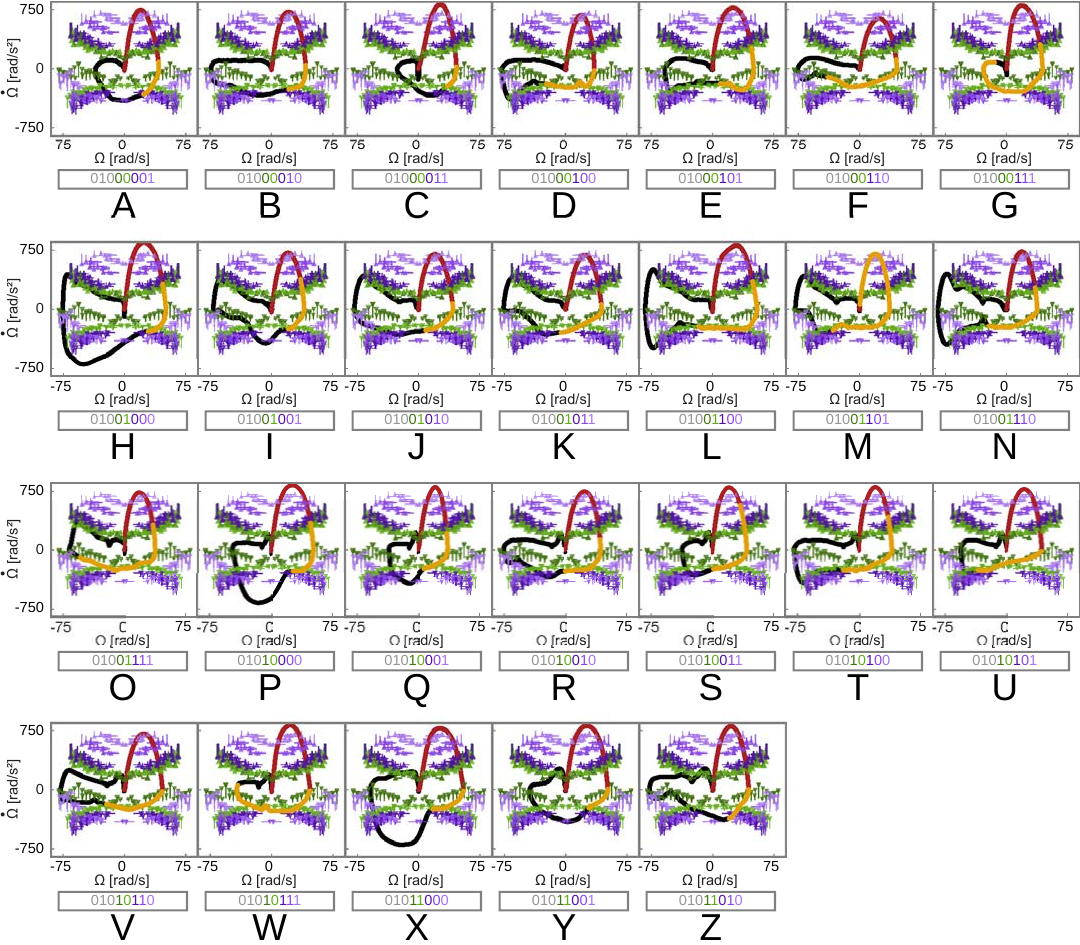}
    \caption{\textbf{Library of selected writing orbits.}
        All 26 selected orbits of system II encode the uppercase Latin character (A–Z) using the ASCII encoding standard~\cite{ASCII}. Although ASCII requires eight bits, the first three bits are identical for all uppercase characters; hence we write the remaining five bits. Below each orbit, the full bit string is shown, with colors labeling the mbits, and gray indicating the three shared bits. The orbits are color-coded, with red indicating erase phase one; gold, erase phase two; and black the writing phase.}
    \label{fig:alphabet}
\end{figure*}

\end{document}